\pdfoutput=1

\documentclass[12pt,a4paper]{article}

\usepackage{ifthen} 
\newboolean{pdflatex}
\setboolean{pdflatex}{true} 

\newboolean{articletitles}
\setboolean{articletitles}{true} 

\newboolean{uprightparticles}
\setboolean{uprightparticles}{false} 

\newboolean{inbibliography}
\setboolean{inbibliography}{false} 

\def\paperauthors{LHCb collaboration} 
\def\paperasciititle{Measurement of Z to tau-pair production in proton-proton collisions at sqrt(s) = 8TeV} 
\def\papertitle{Measurement of \Ztautau production in proton-proton collisions at \sqs = 8\tev} 
\def\paperkeywords{{High Energy Physics}, {LHCb}} 
\def\papercopyright{\the\year\ CERN for the benefit of the LHCb collaboration} 
\def\paperlicence{CC-BY-4.0 licence}
\def\paperlicenceurl{https://creativecommons.org/licenses/by/4.0/}

\usepackage[top=1in, bottom=1.25in, left=1in, right=1in]{geometry}

\usepackage{microtype}
\usepackage{lineno}  
\usepackage{xspace} 
\usepackage{caption} 

\usepackage{graphicx}  
\usepackage{color}
\usepackage{colortbl}
\graphicspath{{./figs/}} 
\DeclareGraphicsExtensions{.pdf,.PDF,png,.PNG}

\usepackage{amsmath} 
\usepackage{amssymb}
\usepackage{amsfonts}
\usepackage{upgreek} 

\newcommand*\patchAmsMathEnvironmentForLineno[1]{%
\expandafter\let\csname old#1\expandafter\endcsname\csname #1\endcsname
\expandafter\let\csname oldend#1\expandafter\endcsname\csname
end#1\endcsname
 \renewenvironment{#1}%
   {\linenomath\csname old#1\endcsname}%
   {\csname oldend#1\endcsname\endlinenomath}%
}
\newcommand*\patchBothAmsMathEnvironmentsForLineno[1]{%
  \patchAmsMathEnvironmentForLineno{#1}%
  \patchAmsMathEnvironmentForLineno{#1*}%
}
\AtBeginDocument{%
\patchBothAmsMathEnvironmentsForLineno{equation}%
\patchBothAmsMathEnvironmentsForLineno{align}%
\patchBothAmsMathEnvironmentsForLineno{flalign}%
\patchBothAmsMathEnvironmentsForLineno{alignat}%
\patchBothAmsMathEnvironmentsForLineno{gather}%
\patchBothAmsMathEnvironmentsForLineno{multline}%
\patchBothAmsMathEnvironmentsForLineno{eqnarray}%
}

\usepackage{hyperxmp}

\usepackage[pdftex,
            pdfauthor={\paperauthors},
            pdftitle={\paperasciititle},
            pdfkeywords={\paperkeywords},
            pdfcopyright={Copyright (C) \papercopyright},
            pdflicenseurl={\paperlicenceurl}]{hyperref}

\usepackage[all]{hypcap} 

\usepackage{xspace} 
\usepackage{upgreek}

\def\lhcb {\mbox{LHCb}\xspace}

\def\presh  {PS\xspace}
\def\ecal   {ECAL\xspace}
\def\hcal   {HCAL\xspace}
\def\MagUp {\mbox{\em Mag\kern -0.05em Up}\xspace}

\ifthenelse{\boolean{uprightparticles}}%
{
 
 \def\Pgamma      {\ensuremath{\upgamma}\xspace}

 \def\Pmu         {\ensuremath{\upmu}\xspace}                 
 \def\Pnu         {\ensuremath{\upnu}\xspace}                 
                  
 \def\Ppi         {\ensuremath{\uppi}\xspace}

 \def\Ptau        {\ensuremath{\uptau}\xspace}

 \def\Ppsi        {\ensuremath{\uppsi}\xspace}

 \def\PDelta      {\ensuremath{\Delta}\xspace}                 
 \def\PXi      {\ensuremath{\Xi}\xspace}                 
 \def\PLambda      {\ensuremath{\Lambda}\xspace}                 
 \def\PSigma      {\ensuremath{\Sigma}\xspace}                 
 \def\POmega      {\ensuremath{\Omega}\xspace}                 
 \def\PUpsilon      {\ensuremath{\Upsilon}\xspace}                 
 
 \def\PB      {\ensuremath{\mathrm{B}}\xspace}                 
                  
 \def\PD      {\ensuremath{\mathrm{D}}\xspace}

 \def\PJ      {\ensuremath{\mathrm{J}}\xspace}                 
 \def\PK      {\ensuremath{\mathrm{K}}\xspace}

 \def\PW      {\ensuremath{\mathrm{W}}\xspace}

 \def\PZ      {\ensuremath{\mathrm{Z}}\xspace}                 
                  
 \def\Pb      {\ensuremath{\mathrm{b}}\xspace}                 
 \def\Pc      {\ensuremath{\mathrm{c}}\xspace}                 
                  
 \def\Pe      {\ensuremath{\mathrm{e}}\xspace}

 \def\Pi      {\ensuremath{\mathrm{i}}\xspace}

 \def\Pt      {\ensuremath{\mathrm{t}}\xspace}

}
{
 
 \def\Pgamma      {\ensuremath{\gamma}\xspace}

 \def\Pmu         {\ensuremath{\mu}\xspace}                 
 \def\Pnu         {\ensuremath{\nu}\xspace}                 
                  
 \def\Ppi         {\ensuremath{\pi}\xspace}

 \def\Ptau        {\ensuremath{\tau}\xspace}

 \def\Ppsi        {\ensuremath{\psi}\xspace}                 
                  
 \mathchardef\PDelta="7101
 \mathchardef\PXi="7104
 \mathchardef\PLambda="7103
 \mathchardef\PSigma="7106
 \mathchardef\POmega="710A
 \mathchardef\PUpsilon="7107
                  
 \def\PB      {\ensuremath{B}\xspace}                 
                  
 \def\PD      {\ensuremath{D}\xspace}

 \def\PJ      {\ensuremath{J}\xspace}                 
 \def\PK      {\ensuremath{K}\xspace}

 \def\PW      {\ensuremath{W}\xspace}

 \def\PZ      {\ensuremath{Z}\xspace}                 
                  
 \def\Pb      {\ensuremath{b}\xspace}                 
 \def\Pc      {\ensuremath{c}\xspace}                 
                  
 \def\Pe      {\ensuremath{e}\xspace}

 \def\Pi      {\ensuremath{i}\xspace}

 \def\Pt      {\ensuremath{t}\xspace}

}

\makeatletter
\ifcase \@ptsize \relax
  \newcommand{\miniscule}{\@setfontsize\miniscule{4}{5}}
\or
  \newcommand{\miniscule}{\@setfontsize\miniscule{5}{6}}
\or
  \newcommand{\miniscule}{\@setfontsize\miniscule{5}{6}}
\fi
\makeatother

\DeclareRobustCommand{\optbar}[1]{\shortstack{{\miniscule (\rule[.5ex]{1.25em}{.18mm})}
  \\ [-.7ex] $#1$}}

\def\electron   {{\ensuremath{\Pe}}\xspace}

\def\epem       {{\ensuremath{\Pe^+\Pe^-}}\xspace}

\def\muon       {{\ensuremath{\Pmu}}\xspace}

\def\mumu       {{\ensuremath{\Pmu^+\Pmu^-}}\xspace}

\def\tauon      {{\ensuremath{\Ptau}}\xspace}
\def\taup       {{\ensuremath{\Ptau^+}}\xspace}

\def\tautau     {{\ensuremath{\Ptau^+\Ptau^-}}\xspace}

\def\neu        {{\ensuremath{\Pnu}}\xspace}

\def\neut       {{\ensuremath{\neu_\tau}}\xspace}

\def\g      {{\ensuremath{\Pgamma}}\xspace}

\def\W      {{\ensuremath{\PW}}\xspace}

\def\Z      {{\ensuremath{\PZ}}\xspace}

\def\cquark    {{\ensuremath{\Pc}}\xspace}

\def\bquark    {{\ensuremath{\Pb}}\xspace}
\def\bquarkbar {{\ensuremath{\overline \bquark}}\xspace}
\def\bbbar     {{\ensuremath{\bquark\bquarkbar}}\xspace}
\def\tquark    {{\ensuremath{\Pt}}\xspace}
\def\tquarkbar {{\ensuremath{\overline \tquark}}\xspace}
\def\ttbar     {{\ensuremath{\tquark\tquarkbar}}\xspace}

\def\pion   {{\ensuremath{\Ppi}}\xspace}
\def\piz    {{\ensuremath{\pion^0}}\xspace}

\def\pip    {{\ensuremath{\pion^+}}\xspace}

\def\kaon    {{\ensuremath{\PK}}\xspace}
  \def\Kbar    {{\kern 0.2em\overline{\kern -0.2em \PK}{}}\xspace}

\def\KorKbar    {\kern 0.18em\optbar{\kern -0.18em K}{}\xspace}

\def\Kp      {{\ensuremath{\kaon^+}}\xspace}
\def\Km      {{\ensuremath{\kaon^-}}\xspace}

  \def\Dbar    {{\kern 0.2em\overline{\kern -0.2em \PD}{}}\xspace}
\def\D       {{\ensuremath{\PD}}\xspace}

\def\DorDbar    {\kern 0.18em\optbar{\kern -0.18em D}{}\xspace}
\def\Dz      {{\ensuremath{\D^0}}\xspace}

\def\Dstarp  {{\ensuremath{\D^{*+}}}\xspace}
\def\Dstarm  {{\ensuremath{\D^{*-}}}\xspace}

\def\B       {{\ensuremath{\PB}}\xspace}
\def\Bbar    {{\ensuremath{\kern 0.18em\overline{\kern -0.18em \PB}{}}}\xspace}

\def\BorBbar    {\kern 0.18em\optbar{\kern -0.18em B}{}\xspace}
\def\Bz      {{\ensuremath{\B^0}}\xspace}

\def\Bu      {{\ensuremath{\B^+}}\xspace}

\def\Bp      {{\ensuremath{\Bu}}\xspace}

\def\jpsi     {{\ensuremath{{\PJ\mskip -3mu/\mskip -2mu\Ppsi\mskip 2mu}}}\xspace}

  \def\Y#1S{\ensuremath{\PUpsilon{(#1S)}}\xspace}

\def\Lbar        {{\ensuremath{\kern 0.1em\overline{\kern -0.1em\PLambda}}}\xspace}
\def\LorLbar    {\kern 0.18em\optbar{\kern -0.18em \PLambda}{}\xspace}

\def\BF         {{\ensuremath{\mathcal{B}}}\xspace}

\def\BR         {\BF}
\newcommand{\decay}[2]{\ensuremath{#1\!\to #2}\xspace}         

\def\to                 {\ensuremath{\rightarrow}\xspace}

\newcommand{\erec}{{\ensuremath{\varepsilon_{\mathrm{ rec/det}}}}\xspace}
\newcommand{\esel}{{\ensuremath{\varepsilon_{\mathrm{ sel/rec}}}}\xspace}

\def\AT#1     {\ensuremath{A_{\mathrm{T}}^{#1}}\xspace}           

\def\C#1      {\ensuremath{\mathcal{C}_{#1}}\xspace}                       
\def\Cp#1     {\ensuremath{\mathcal{C}_{#1}^{'}}\xspace}                    
\def\Ceff#1   {\ensuremath{\mathcal{C}_{#1}^{\mathrm{(eff)}}}\xspace}        
\def\Cpeff#1  {\ensuremath{\mathcal{C}_{#1}^{'\mathrm{(eff)}}}\xspace}       
\def\Ope#1    {\ensuremath{\mathcal{O}_{#1}}\xspace}                       
\def\Opep#1   {\ensuremath{\mathcal{O}_{#1}^{'}}\xspace}                    

\newcommand{\tev}{\ifthenelse{\boolean{inbibliography}}{\ensuremath{~T\kern -0.05em eV}}{\ensuremath{\mathrm{\,Te\kern -0.1em V}}}\xspace}
\newcommand{\gev}{\ensuremath{\mathrm{\,Ge\kern -0.1em V}}\xspace}
\newcommand{\mev}{\ensuremath{\mathrm{\,Me\kern -0.1em V}}\xspace}
\newcommand{\kev}{\ensuremath{\mathrm{\,ke\kern -0.1em V}}\xspace}
\newcommand{\ev}{\ensuremath{\mathrm{\,e\kern -0.1em V}}\xspace}
\newcommand{\gevc}{\ensuremath{{\mathrm{\,Ge\kern -0.1em V\!/}c}}\xspace}
\newcommand{\mevc}{\ensuremath{{\mathrm{\,Me\kern -0.1em V\!/}c}}\xspace}
\newcommand{\gevcc}{\ensuremath{{\mathrm{\,Ge\kern -0.1em V\!/}c^2}}\xspace}
\newcommand{\gevgevcccc}{\ensuremath{{\mathrm{\,Ge\kern -0.1em V^2\!/}c^4}}\xspace}
\newcommand{\mevcc}{\ensuremath{{\mathrm{\,Me\kern -0.1em V\!/}c^2}}\xspace}

\def\mum  {\ensuremath{{\,\upmu\mathrm{m}}}\xspace}

\def\pb {\ensuremath{\mathrm{ \,pb}}\xspace}
\def\invpb {\ensuremath{\mbox{\,pb}^{-1}}\xspace}

\def\invfb   {\ensuremath{\mbox{\,fb}^{-1}}\xspace}

\def\fs   {\ensuremath{\mathrm{ \,fs}}\xspace}

\newcommand{\chisq}{\ensuremath{\chi^2}\xspace}

\def\gsim{{~\raise.15em\hbox{$>$}\kern-.85em
          \lower.35em\hbox{$\sim$}~}\xspace}
\def\lsim{{~\raise.15em\hbox{$<$}\kern-.85em
          \lower.35em\hbox{$\sim$}~}\xspace}

\def\sqs   {\ensuremath{\protect\sqrt{s}}\xspace}

\def\pt         {\ensuremath{p_{\mathrm{ T}}}\xspace}
\def\ptot       {\ensuremath{p}\xspace}

\newcommand{\lum} {\ensuremath{\mathcal{L}}\xspace}

\def\evtgen     {\mbox{\textsc{EvtGen}}\xspace}
\def\fewz       {\mbox{\textsc{Fewz}}\xspace}

\def\geant      {\mbox{\textsc{Geant4}}\xspace}

\def\photos     {\mbox{\textsc{Photos}}\xspace}

\def\pythia     {\mbox{\textsc{Pythia}}\xspace}

\def\tell1  {TELL1\xspace}
\def\ukl1   {UKL1\xspace}

\newcommand{\ie}{\mbox{\itshape i.e.}\xspace}

\usepackage{cite} 
\usepackage{mciteplus}

\usepackage{mathtools}
\usepackage{booktabs,siunitx}
\usepackage{float} 
\usepackage{subfigure}
\usepackage[xcolor,pdftex,rightbars]{changebar} 
\cbcolor{red}

\hypersetup{
  colorlinks=true,
  citecolor=blue,
  linkcolor=red,
}

\usepackage{cleveref}
\crefname{section}{Sect.}{Sects.}
\crefname{subsection}{Sect.}{Sects.}
\crefname{subsubsection}{Sect.}{Sects.}
\crefname{figure}{Fig.}{Figs.}
\crefname{table}{Table}{Tables}

\usepackage{setspace}

\newcommand\To{\ensuremath{\rightarrow\;}} 

\newcommand{\taue}{\ensuremath{\tau_e}\xspace}
\newcommand{\taumu}{\ensuremath{\tau_\mu}\xspace}

\newcommand{\tauh}[1]{\ensuremath{\tau_{h#1}}\xspace}

\newcommand{\ditaux}[2]{\ensuremath{\tau_{#1}\tau_{#2}}\xspace}
\newcommand{\ditauee}{\ditaux{e}{e}}
\newcommand{\ditaumue}{\ditaux{\mu}{e}}
\newcommand{\ditaumumu}{\ditaux{\mu}{\mu}}
\newcommand{\ditaueh}[1]{\ditaux{e}{h#1}}
\newcommand{\ditaumuh}[1]{\ditaux{\mu}{h#1}}

\newcommand{\dee}{\ditauee{}\xspace}
\newcommand{\dmue}{\ditaumue{}\xspace}
\newcommand{\dmumu}{\ditaumumu{}\xspace}
\newcommand{\deh}[1]{\ditaueh{#1}}
\newcommand{\dmuh}[1]{\ditaumuh{#1}}

\newcommand{\Zll}{\decay{\Z}{l^+l^-}}

\newcommand{\Ztautau}{\decay{\Z}{\tautau}}

\newcommand{\Zmumu}{\decay{\Z}{\muon^+\muon^-}}

\newcommand{\Zee}{\decay{\Z}{\electron^+\electron^-}}

\newcommand{\zbb}{\decay{\Z}{\bbbar}}
\newcommand{\wjet}{\W{}+\text{jet}\xspace}
\newcommand{\zjet}{\Z{}+\text{jet}\xspace}
\newcommand{\Vj}{{\ensuremath{V\!j}}\xspace}
\newcommand{\VV}{{\ensuremath{VV}}\xspace} 

\newcommand{\ppZtautau}{\decay{pp}{\Ztautau}\xspace}

\newcommand{\ppZmumu}{\decay{pp}{\Zmumu}\xspace}

\newcommand{\ppZee}{\decay{pp}{\Zee}\xspace}

\newcommand{\mass}{\ensuremath{m}\xspace}
\newcommand{\Eta}{\ensuremath{\eta}\xspace}

\newcommand{\Retaphi}{\ensuremath{R_{\eta\phi}}\xspace}

\newcommand{\mcorr}{\ensuremath{m_\text{corr}}\xspace}

\newcommand{\ipthat}{\ensuremath{\hat{I}_{p_{\rm T}}}\xspace}
\newcommand{\apt}{\ensuremath{A_{p_{\rm T}}}\xspace}

\newcommand{\deltapt}{$\pt{(\tau_1)}-\pt{(\tau_2)}$\xspace}

\newcommand{\nss}[1]{\ensuremath{N^\text{SS}_\text{#1}}\xspace}

\newcommand{\eff}[1]{\ensuremath{\varepsilon_{#1}}\xspace}
\newcommand{\eacc}{\ensuremath{\mathcal{A}}\xspace}

\renewcommand{\erec}{\eff{\text{rec}}} 
\renewcommand{\esel}{\eff{\text{sel}}} 
\newcommand{\nobs}{\ensuremath{N_\text{obs}}\xspace}

\newcommand{\dynnlo}{\textsc{Dynnlo}\xspace}
\newcommand{\nlo}{next-to-leading order\xspace}

\begin{document}

\renewcommand{\thefootnote}{\fnsymbol{footnote}}
\setcounter{footnote}{1}


\begin{titlepage}
\pagenumbering{roman}

\vspace*{-1.5cm}
\centerline{\large EUROPEAN ORGANIZATION FOR NUCLEAR RESEARCH (CERN)}
\vspace*{1.5cm}
\noindent
\begin{tabular*}{\linewidth}{lc@{\extracolsep{\fill}}r@{\extracolsep{0pt}}}
\ifthenelse{\boolean{pdflatex}}
{\vspace*{-1.5cm}\mbox{\!\!\!\includegraphics[width=.14\textwidth]{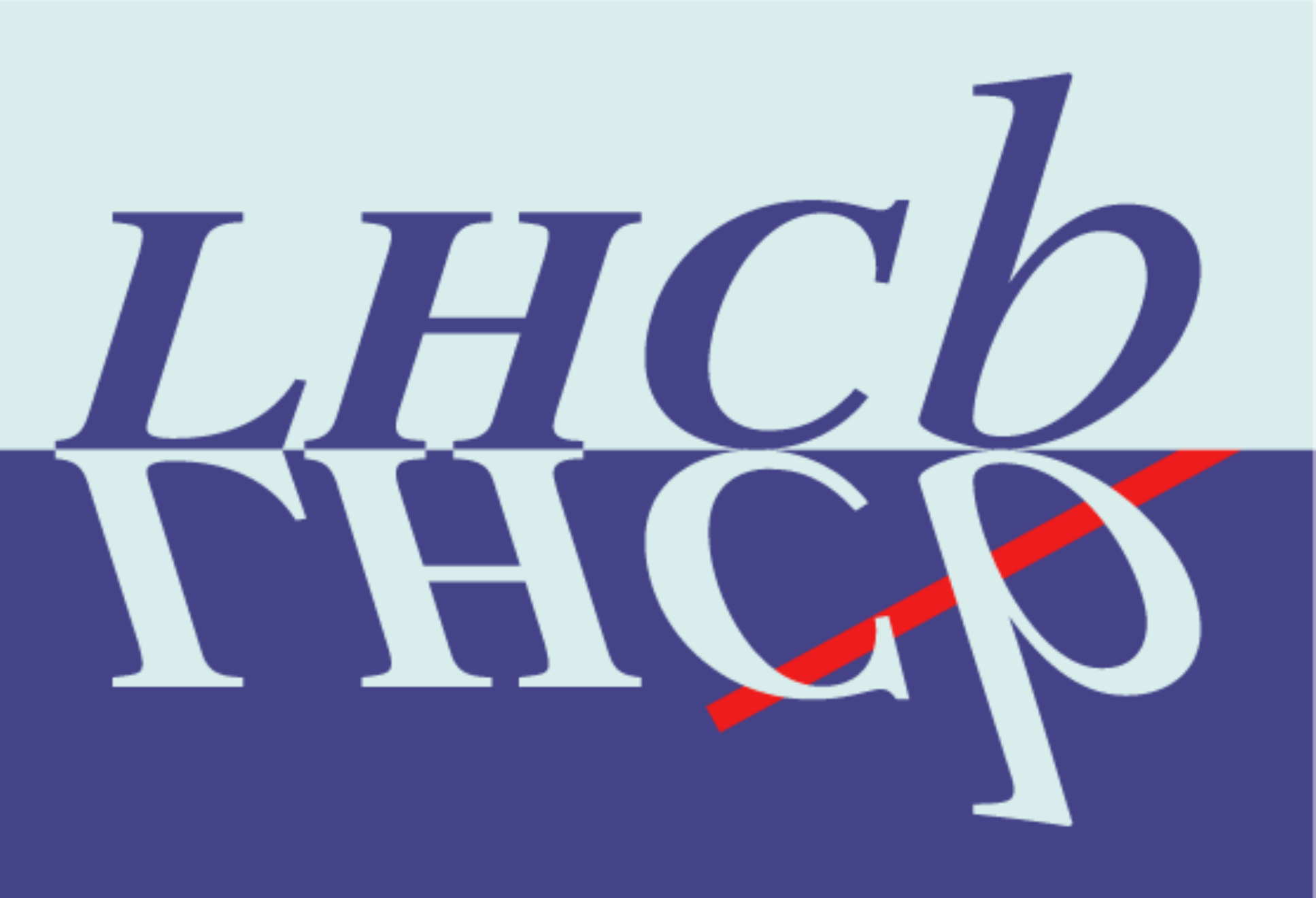}} & &}%
{\vspace*{-1.2cm}\mbox{\!\!\!\includegraphics[width=.12\textwidth]{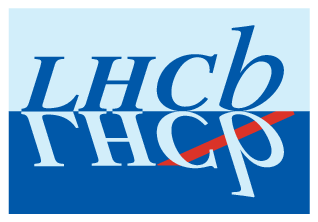}} & &}%
\\
 & & CERN-EP-2018-147 \\  
 & & LHCb-PAPER-2018-016 \\  
 & & September 27, 2018 \\
\end{tabular*}

\vspace*{2.5cm}

{\normalfont\bfseries\boldmath\huge
\begin{center}
  \papertitle 
\end{center}
}

\vspace*{2.0cm}

\begin{center}
\paperauthors\footnote{Authors are listed at the end of this paper.}
\end{center}

\vspace{\fill}

\begin{abstract}
  \noindent
  A measurement of \Ztautau production cross-section is presented using data, 
  corresponding to an integrated luminosity of 2\invfb, 
  from $pp$ collisions at \sqs= 8\tev collected by the \lhcb{} experiment. 
  The \tautau candidates are reconstructed in final states 
  with the first tau lepton decaying leptonically,
  and the second decaying either leptonically or to one or three charged hadrons.
  The production cross-section is measured for \Z bosons 
  with invariant mass between 60 and 120\gevcc,
  which decay to tau leptons with transverse momenta greater than 20\gevc
  and pseudorapidities between 2.0 and 4.5.
%
%
  The cross-section is determined to be
  $\sigma_{\ppZtautau} = 95.8 \pm 2.1 \pm 4.6 \pm 0.2 \pm 1.1 \pb$,
  where the first uncertainty is statistical, the second is systematic, 
  the third is due to the LHC beam energy uncertainty, 
  and the fourth to the integrated luminosity uncertainty.
  This result is compatible with NNLO Standard model predictions.
  The ratio of the cross-sections for \Ztautau to \Zmumu (\Zee),
  determined to be $1.01 \pm 0.05$ ($1.02 \pm 0.06$),
  is consistent with the lepton-universality hypothesis in \Z decays.
  
\end{abstract}

\vspace*{1.5cm}

\begin{center}
  Published in JHEP 09 (2018) 159
\end{center}

\vspace{\fill}

{\footnotesize 
\centerline{\copyright~\papercopyright. \href{\paperlicenceurl}{\paperlicence}.}}
\vspace*{2mm}

\end{titlepage}


\newpage
\setcounter{page}{2}
\mbox{~}
%
%
%
%

\cleardoublepage


\renewcommand{\thefootnote}{\arabic{footnote}}
\setcounter{footnote}{0}


\pagestyle{plain} 
\setcounter{page}{1}
\pagenumbering{arabic}


\section{Introduction}

The measurement of the production cross-section for
a \Z boson\footnote{\Z refers to \Z/\g{$^\star$}, \ie includes contributions
from the virtual photon production and interference.} 
using different decay modes in proton-proton ($pp$) collisions,
$\sigma_{\decay{pp}{\decay{Z}{f\bar{f}}}}$,
is an important verification of Standard Model (SM) predictions. 
The ratio of the \Ztautau production cross-sections to other leptonic decay modes
provides a test of lepton universality (LU).
The LEP experiments have performed high accuracy tests of LU at the \Z pole, 
with a precision better than 1\%~\cite{ALEPH:2005ab}. 
Consequently, the observation in proton-proton collisions of any apparent 
deviation from LU in \Z decays would be an evidence of new phenomena 
producing final-state leptons, 
like in the theoretical context of mSUGRA~\cite{Heinemann:2003wv},
constrained NMSSM~\cite{Ellwanger:2010es},
Randall-Sundrum models~\cite{Perez:2008ee,Casagrande:2008hr},
or lepton-violating decays of Higgs-like bosons~\cite{Harnik:2012pb,
Blankenburg:2012ex,DiazCruz:1999xe,Goudelis:2011un,Arhrib:2012ax}.

This analysis extends the \lhcb results obtained with $pp$ collisions at
$\sqs=7\tev$~\cite{LHCb-PAPER-2012-029} to $\sqs=8\tev$.
The cross-section is measured for leptons from the \Z decay with
transverse momentum (\pt) above 20\gevc
and a \Z invariant mass between 60 and 120\gevcc,
as for the previously published \Zmumu and \Zee cross-sections
~\cite{LHCb-PAPER-2015-049, LHCb-PAPER-2015-003}.
The cross-section measurements in the pseudorapidity range $2.0<\eta<4.5$
covered by the \lhcb experiment are complementary to those with
the central detectors ATLAS \cite{Limosani:2016wnk} and CMS \cite{Chatrchyan:2011nv}.

In the present analysis, the reconstruction of the tau-pair candidates
is performed in both leptonic and hadronic decay modes of the tau,
requiring at least one leptonic mode for the tau-pair candidate.
The reconstruction of high-\pt tau leptons in the 3-prong decay mode is
performed for the first time in \lhcb.


\section{Detector and datasets}

The \lhcb detector~\cite{Alves:2008zz,LHCb-DP-2014-002} is a single-arm forward spectrometer
designed for the study of particles containing \bquark or \cquark quarks. 
The detector includes a high-precision tracking system
consisting of a silicon-strip vertex detector surrounding the $pp$
interaction region, a large-area silicon-strip detector located
upstream of a dipole magnet with a bending power of $4{\mathrm{\,Tm}}$, 
and three stations of silicon-strip detectors and straw
drift tubes placed downstream of the magnet.
The tracking system provides a measurement of momentum of charged particles with
a relative uncertainty that varies from 0.5\% at low momentum to 1.0\% at 200\gevc.
The minimum distance of a track to a primary vertex (PV), the impact parameter (IP), 
is measured with a resolution of $(15+29/\pt)\mum$,
where \pt is the component of the momentum transverse to the beam, in\,\gevc.
Photons, electrons and hadrons are identified by a calorimeter system consisting of
scintillating-pad (SPD) and preshower detectors (\presh), an electromagnetic
calorimeter (\ecal) and a hadronic calorimeter (\hcal). 
Muons are identified by a system composed of five stations of alternating layers 
of iron and multiwire proportional chambers.

The online event selection is performed by a trigger, 
which consists of a hardware stage, 
based on information from the calorimeter and muon systems, 
followed by a software stage, which applies a full event reconstruction.
The hardware trigger imposes a global event cut (GEC) 
requiring the hit multiplicity in the SPD to be less than 600,
to prevent events with high occupancy from dominating the 
processing time in the software trigger.

This analysis uses $pp$ collisions at $\sqs = 8\tev$
corresponding to a total integrated luminosity of
$\lum = (1976 \pm 23) \invpb$~\cite{LHCb-PAPER-2014-047}.
Simulated data samples are used to study the event selection,
determine efficiencies, and estimate systematic uncertainties.
In the simulation, $pp$ collisions are generated using
\pythia8~\cite{Sjostrand:2007gs,*Sjostrand:2006za} 
with a specific \lhcb configuration~\cite{LHCb-PROC-2010-056},
and parton density functions taken from CTEQ6L~\cite{cteq6l}.
Decays of hadronic particles are described by \evtgen~\cite{Lange:2001uf}, 
in which final-state radiation is generated using \photos~\cite{Golonka:2005pn}. 
The interaction of the generated particles with the detector, and its response,
are implemented using the \geant toolkit~\cite{Allison:2006ve, *Agostinelli:2002hh}
as described in Ref.~\cite{LHCb-PROC-2011-006}.


\vspace*{-5pt}
\section{Event selection}

The \Z boson is reconstructed from \tauon particles decaying into leptonic (muons or electrons)
or hadronic (one or three charged hadrons) final states.
Charged tracks are reconstructed by the tracking system and matched 
with clusters of ECAL/HCAL cells and hits in the muon detector.
Muon candidates are identified by matching tracks to hits in the muon
stations downstream of the calorimeters.
They are required to leave hits in at least three muon stations,
or four muon stations if they have $\pt > 10\gevc$.
Electron candidates must fail the muon identification criteria and
fall within the acceptance of the \presh, \ecal, and \hcal sub-detectors.
On average, 30\% of a material radiation length is crossed by a particle before the bending
magnet, causing a considerable energy loss by bremsstrahlung for electrons and positrons.
Hence, the electron or positron candidate momentum is corrected using a bremsstrahlung photon recovery technique~\cite{brem}.
However, since the ECAL is designed to register particles from heavy-flavour hadron decays,
calorimeter cells with transverse energy above about 10\gev saturate the electronics,
and lead to incomplete electron bremsstrahlung recovery.
A large energy deposit in the \presh, \ecal, but not in \hcal is required,
satisfying $E_\text{\presh} > 50\mev$,
$E_\text{\ecal}/\ptot > 0.1$, and $E_\text{\hcal}/\ptot < 0.05$, 
where \ptot is the reconstructed momentum of the electron candidate, 
after applying the bremsstrahlung photon recovery.
Charged hadrons are required to be within the \hcal acceptance, deposit an energy of
$E_\text{\hcal}/\ptot > 0.05$, and must fail the muon identification criteria.
The pion mass is assigned to all charged hadrons.

The analysis is divided into seven ``streams'', labelled as
\ditaumumu, \ditaumue, \ditaumuh1, \ditaumuh3, \ditauee, \ditaueh1, and \ditaueh3,
where the subscript denotes the final state reconstructed.
Charge-conjugate processes are implied throughout.
The streams are chosen such that at least one \tauon lepton decays leptonically.
The tau-pair candidates are selected by triggers requiring muons or electrons 
with a minimum transverse momentum of 15\gevc.
The trigger efficiency is between 70\% and 85\%,
depending on the number of leptons in the stream.
The final states presented in this analysis account for 58\% of all \Ztautau decays.
In the following, a \tauon candidate corresponds to a single particle
for the \taue, \taumu, and \tauh1 decay channels, 
or a combination of the three hadrons in the case of \tauh3.
A pair of \tauon candidates must be associated to the same PV.
In case where multiple PVs are presented in the event,
the associated PV is defined as that with a smallest change in vertex-fit \chisq
when it is reconstructed with and without the \tauon candidate.

The dominating backgrounds are of QCD origin with one or several jets 
(call ``QCD events'' in the following), as well as electroweak processes, 
mainly \W/\Z+jets (``\Vj'').
The following requirements on the transverse momentum of \tauon decay products
are used to reduce these backgrounds.
For all the streams the triggering lepton must have $\pt > 20\gevc$.
For the processes \ditaumumu, \ditauee, and \ditaumue the second lepton \pt threshold is 5\gevc.
The hadron of the \tauh1 candidates is required to have $\pt > 10\gevc$.
For the \tauh3 decay channel, 
each of the three charged hadrons are selected with $\pt > 1\gevc$,
and at least one must be above 6\gevc.
In addition, the \tauh3 candidates must have a total \pt in excess of 12\gevc,
and an invariant mass in the range 0.7 to 1.5\gevcc.
This leads to the \tauh3 identification efficiency of about 30\%, 
comparable with the value of 35\% found in the context of the 
\decay{\Bz}{\Dstarm\taup\neut} analysis \cite{Aaij:2017uff}.
For all streams, the reconstructed direction of the \tauon candidate
must be in the fiducial geometrical acceptance $2.0<\eta<4.5$.

Additional selection criteria are needed to suppress background processes
due to semileptonic \cquark- or \bquark-hadron decays,
misidentification of hadrons as leptons,
or, especially in the \tauh3 stream, combinations of unrelated particles.

Signal candidates tend to have back-to-back tracks in the plane transverse to 
the beam axis, and a higher invariant mass than the background.
Hence, the tau-pair is required to have an invariant mass above 20\gevcc,
or 30\gevcc for the stream containing \tauh1,\tauh3 candidates.
Additionally, for the dilepton streams \ditaumumu, \ditauee,
the selected mass range is below 60\gevcc,
to avoid the on-shell \Zmumu and \Zee regions.
The absolute difference in azimuthal angle of the two \tauon candidates is required
to be greater than $2.7$ radians.
The above selections are found to be 70 to 80\% efficient,
depending on the analysis stream.

Charged particles in QCD events tend to be associated with jet activity, 
in contrast to signal candidates where they are isolated. 
An isolation variable, \ipthat, is defined as the \pt of the candidate 
divided by the transverse component of the vectorial sum of all track momenta
in a cone surrounding the candidate of radius $\Retaphi=0.5$,
defined in the pseudorapidity-azimuthal angle $(\eta-\phi)$ space.
A fully isolated candidate has $\ipthat=1$, 
while lower values indicate the presence of jet activity.
The selection $\ipthat>0.9$ is applied to all \tauon candidates, 
with an efficiency of more than 64\% for the tau-pair signal
and rejecting about 98\% of QCD events.

The lifetime of the \tauon lepton is used to separate the signal from prompt background.
For the \tauon decay channels with a single charged particle, 
it is not possible to reconstruct a secondary vertex 
and a selection on the particle IP to the associated PV is applied.
The efficiency on the signal from these criteria is in the range 71 to 79\%.

\begin{figure}[!t]
\centering
\includegraphics[width=.6\textwidth]{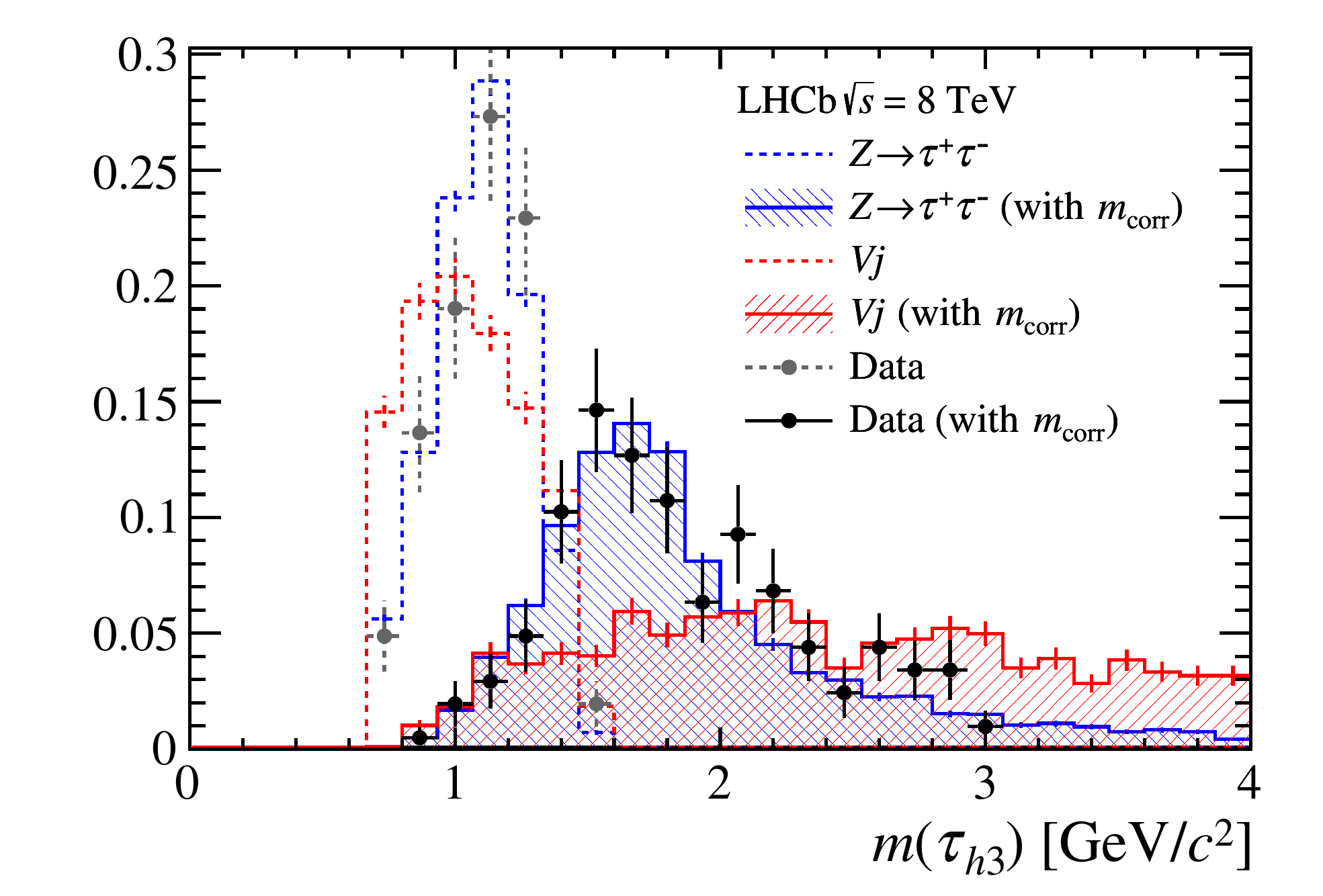}
\caption{\label{fig:mcorr}
  Distributions of invariant (dashed line) and corrected (full line, shaded)
  mass of \tauh3 candidates from the \dmuh3 channel. 
  The yields are normalised to the integrated luminosity of the data.
  The results from data are represented by the black points.
  The error bars represent the statistical uncertainty only.
  The distributions are compared to the signal distributions 
  from simulated \Ztautau (blue) events and the \Vj (red) background.
}
\end{figure}

In the \tauh3 case, a vertex reconstruction is possible:
the maximum distance between the three tracks in the $\eta-\phi$ space
is required to be less than $0.005 \cdot \pt$ 
where \pt is the transverse momentum of \tauh3 in \gevc.
The proper decay time is subsequently estimated from the distance of the 
reconstructed vertex to the associated PV, and the momentum of the candidate, 
taken as an approximation of the \tauon momentum.
A minimum of 60\fs is imposed for this variable, 
efficiently discarding the prompt background
whilst keeping about 77\% of the signal.
For the \tauh3 decay, a correction to the mass is also possible by exploiting 
the direction of flight, recovering part of the momentum lost due to
undetected particles. 
The corrected mass is defined as
\begin{equation} 
\mcorr \equiv \sqrt{m^2 + p^2 \sin^2\theta} + p\sin\theta\,,
\end{equation}
where \mass and \ptot are the invariant mass and momentum computed from the three tracks
and $\theta$ is the angle between the momentum and flight direction of the candidate.
The requirement $\mcorr < 3\gevcc$ reduces the QCD background by about 50\% and the \Vj background by about 60\%, retaining 80\% of the signal.
\Cref{fig:mcorr} shows the mass distributions of \tauh3 candidates before and after correction for data, compared to the distributions of \Ztautau decays and of the \Vj background from simulation.

In the \ditauee and \ditaumumu streams 
an additional background component arises from \Zll decays.
This process produces two muons or two electrons with similar \pt values,
in contrast to signal which tends to have unbalanced \pt
due to the missing momentum from unreconstructed neutrinos and neutral hadrons. 
The \pt asymmetry, \apt, is defined as the absolute \pt difference
of the two candidates divided by their sum. 
For the two leptonic streams \apt is required to be greater than $0.1$.
A particular case is the \ditaumue stream, where background from \Vj processes arises,
with one lepton coming from the jet causing a relatively large \pt imbalance 
with respect to the lepton from the $\W/\Z$ boson.
A suppression by a factor of two of this source of background, 
with a loss of 10\% of the signal is obtained imposing a maximal \apt value of 0.6.
For \tauh1 and \tauh3 the \apt criterion has been found inefficient for background rejection,
hence no such a constraint is imposed to these two decay modes.


\section{Signal and background estimation}

\begin{figure}[!t]
  \newlength{\pw}
  \setlength{\pw}{0.52\textwidth}
  \newcommand{\lw}[1]{\put(-197,138){(#1)}}
  \centering
\mbox{
  \centering
  \subfigure{\includegraphics[width=\pw]{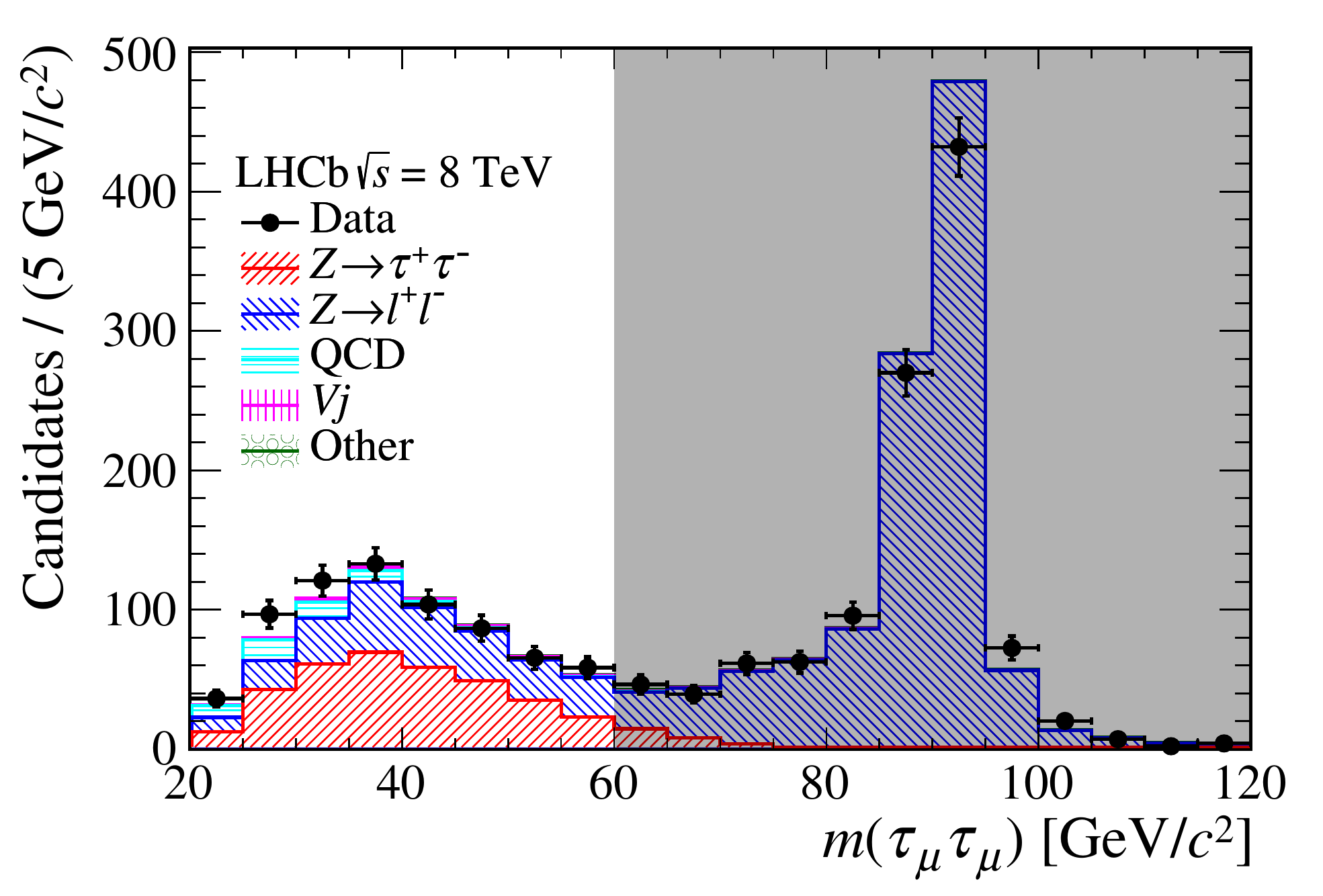} \lw{a} \label{fig:mass_mumu}}\hspace*{-8pt}
  \subfigure{\includegraphics[width=\pw]{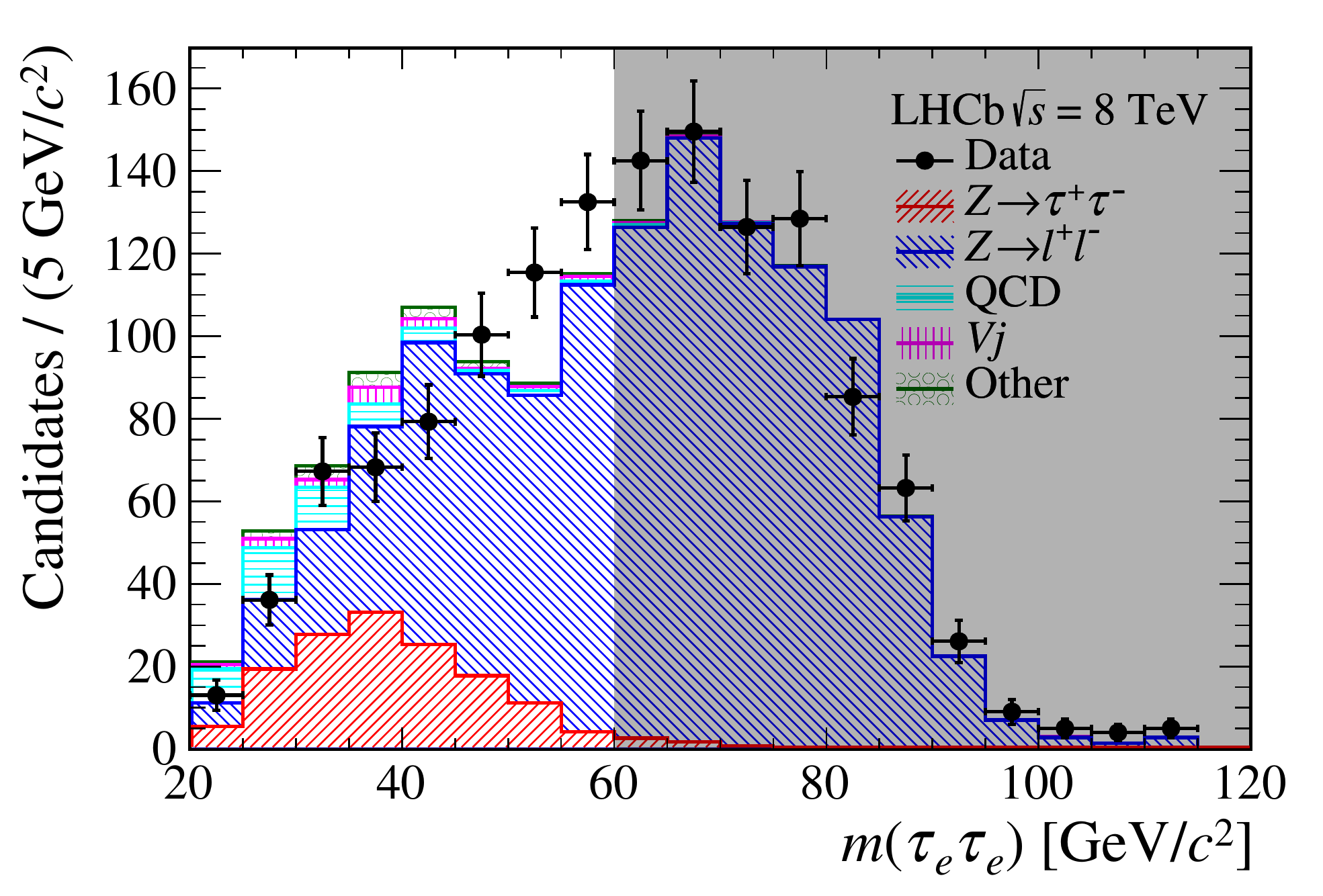} \lw{b}}
}\vspace*{-12pt}
\mbox{
  \centering
  \subfigure{\includegraphics[width=\pw]{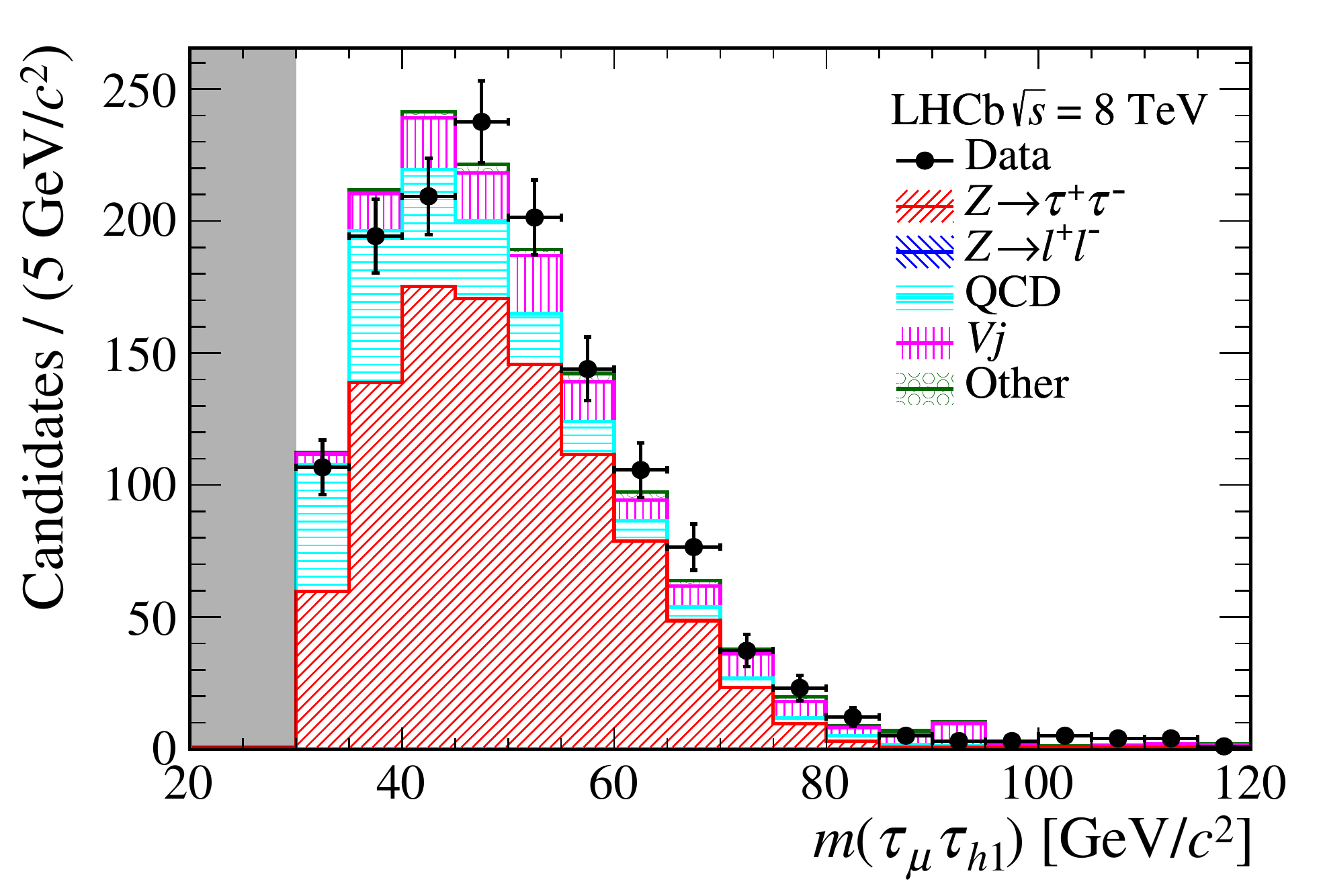} \lw{c}}\hspace*{-8pt}
  \subfigure{\includegraphics[width=\pw]{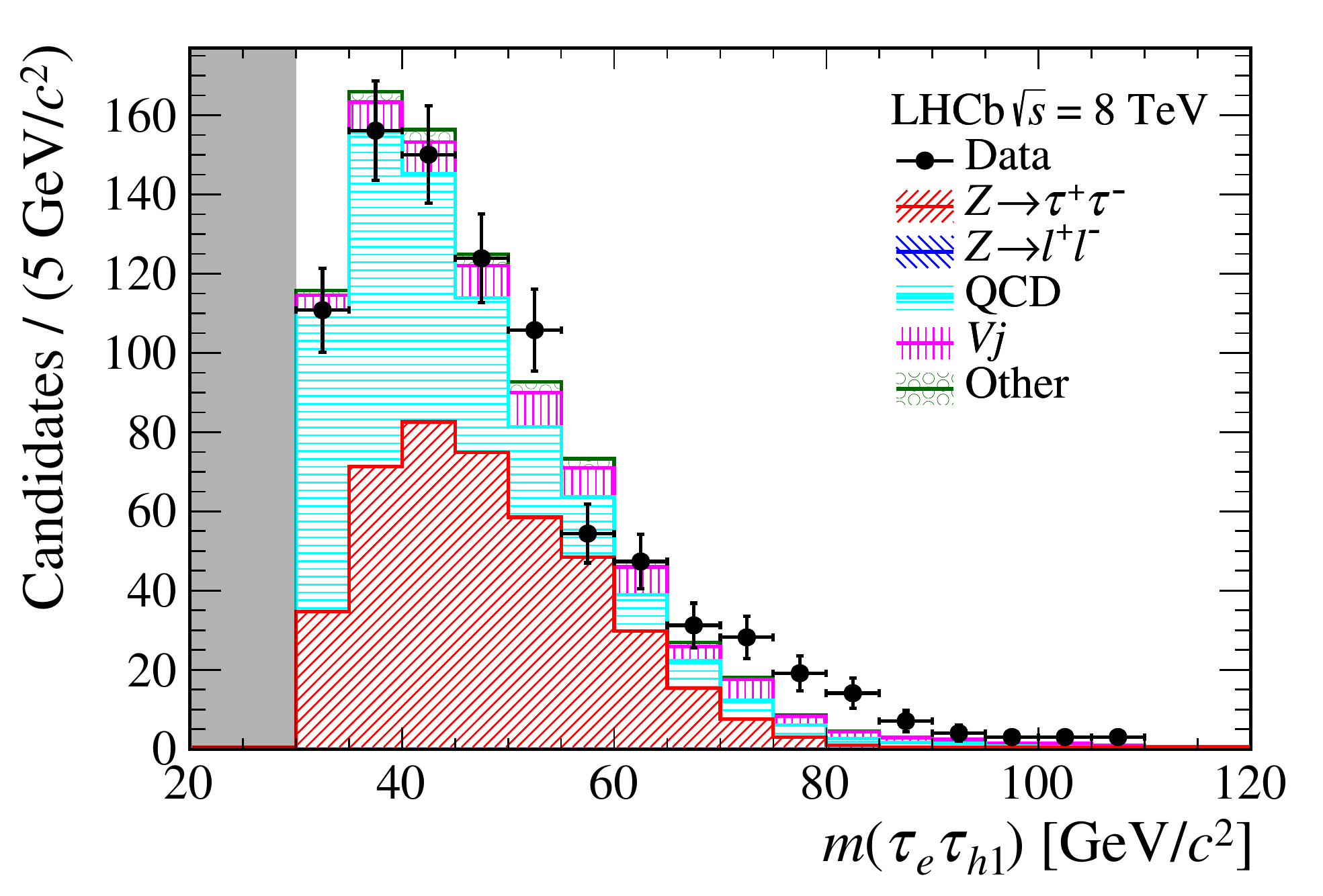} \lw{d}}
}\vspace*{-12pt}
\mbox{
  \centering
  \subfigure{\includegraphics[width=\pw]{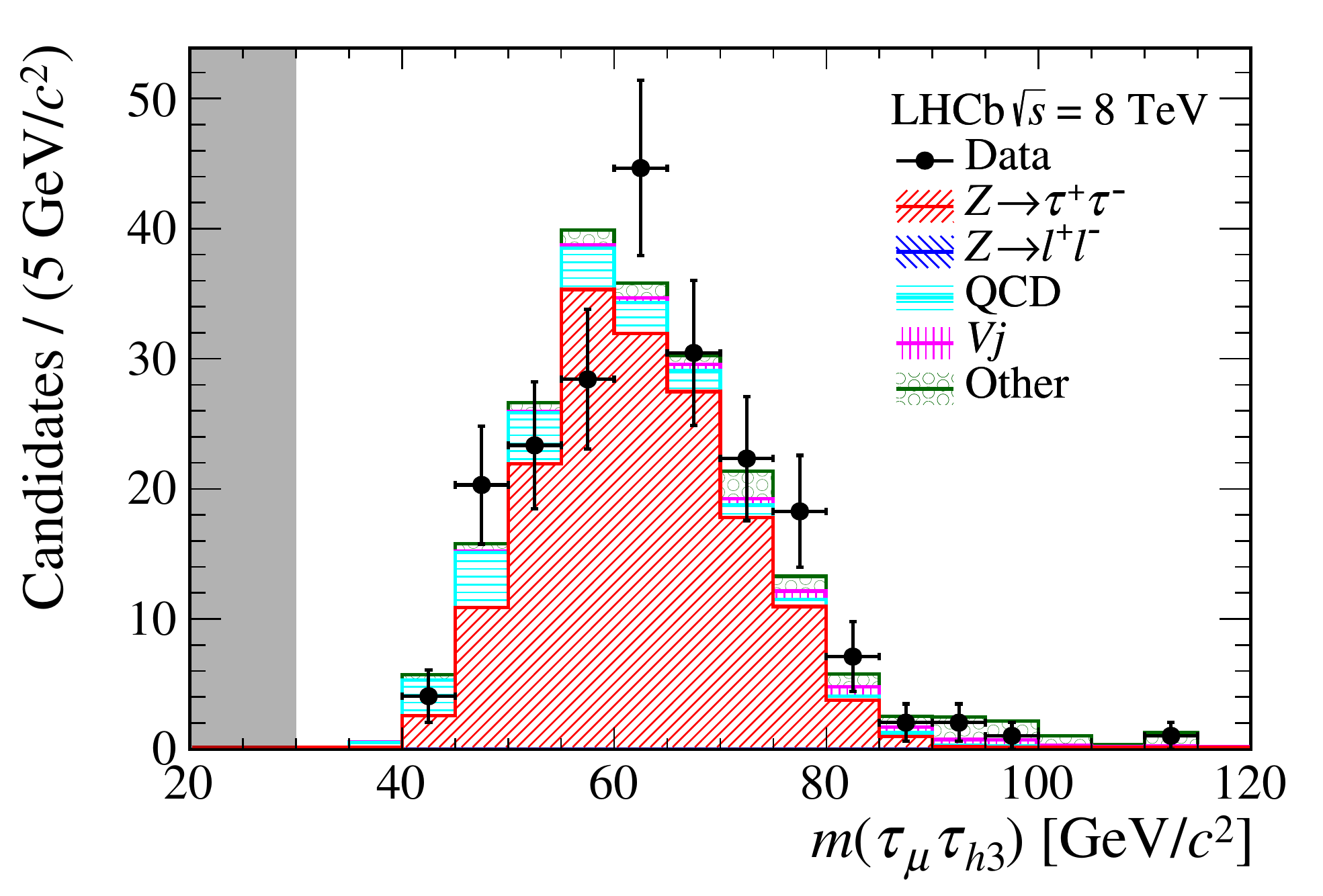} \lw{e}}\hspace*{-8pt}
  \subfigure{\includegraphics[width=\pw]{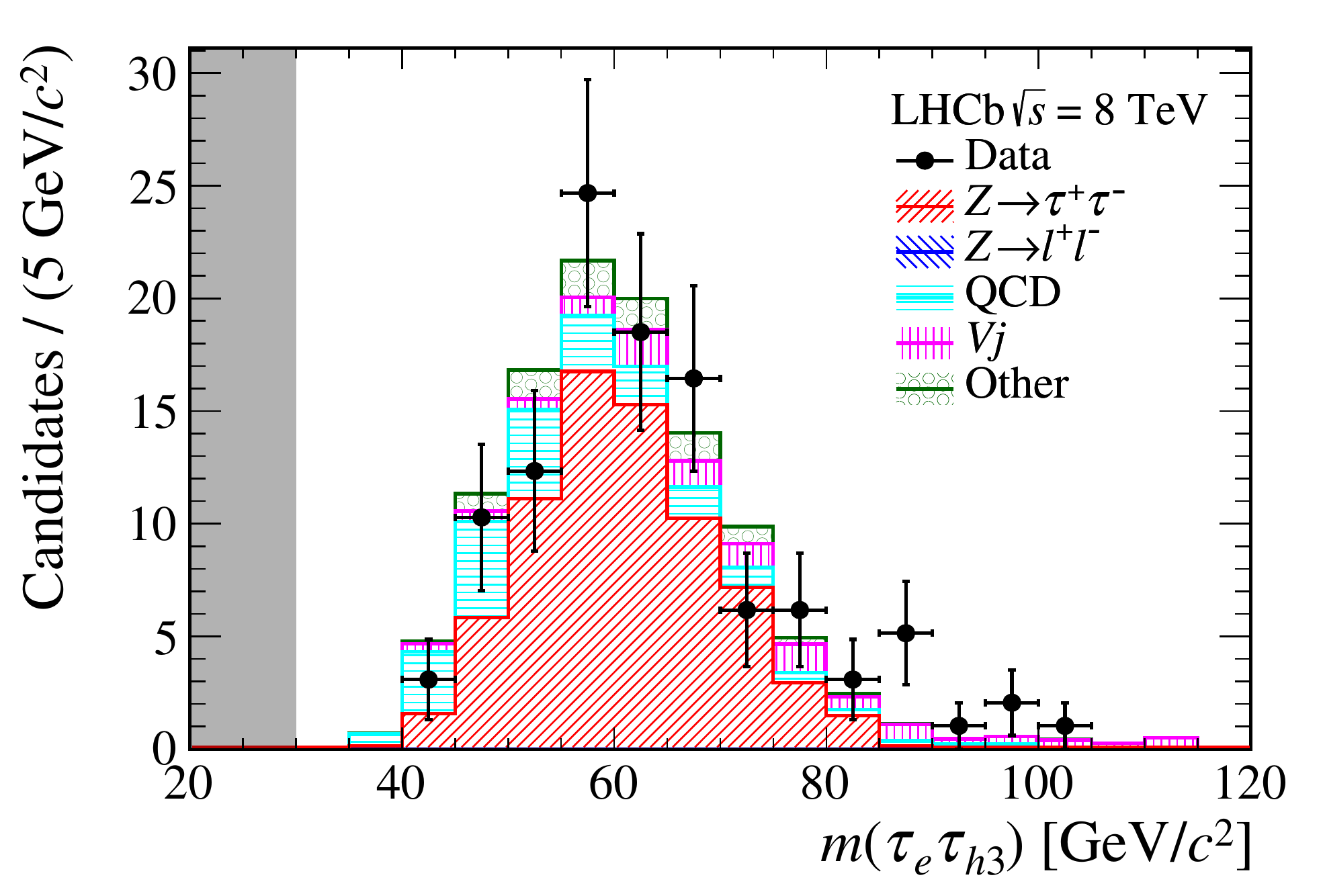} \lw{f}}
}
\makebox[0.5\textwidth][c]{
  \subfigure{\includegraphics[width=\pw]{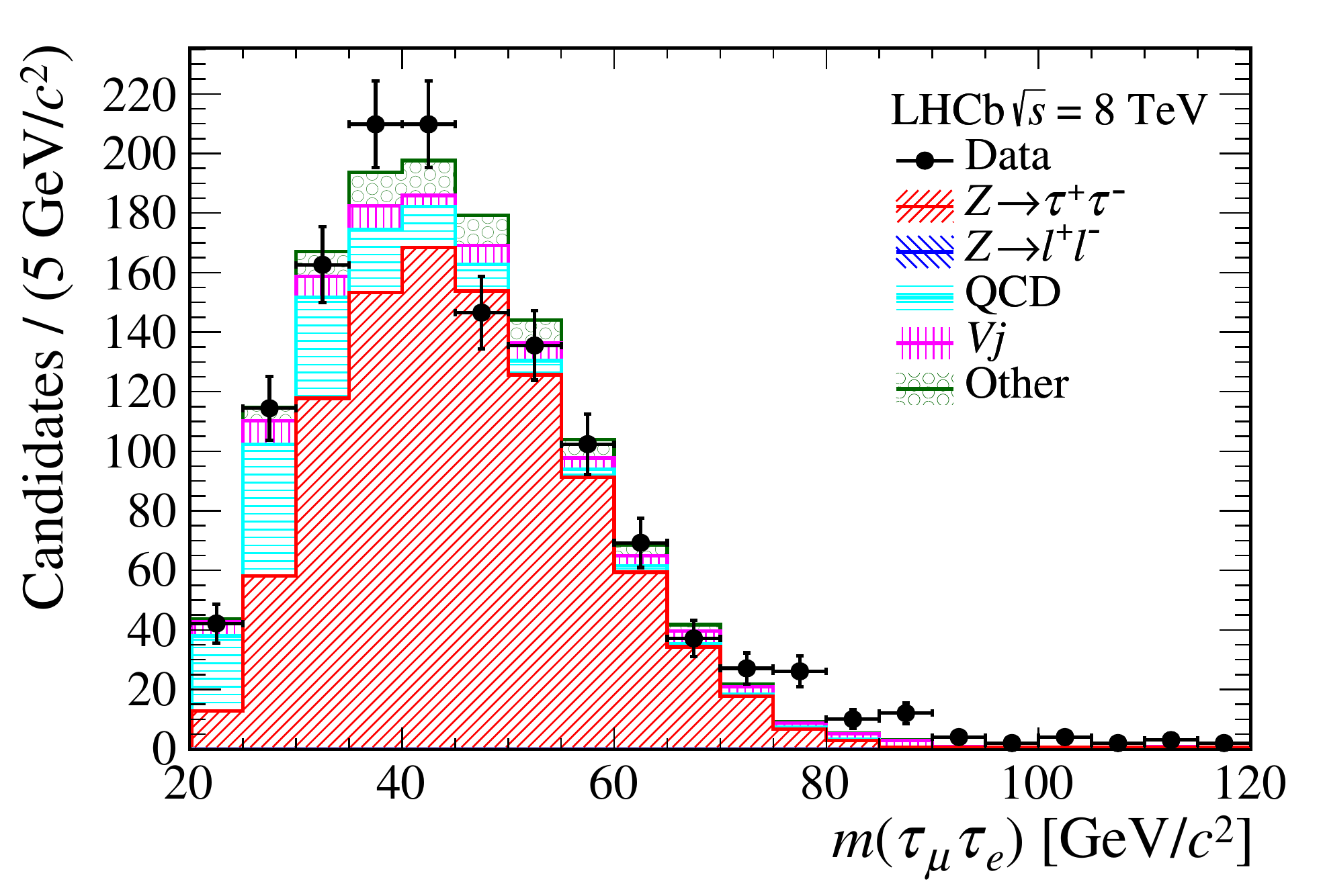} \lw{g}}\hspace*{-8pt} 
}
\hfill
\begin{minipage}[b]{0.46\textwidth}
  \caption{\label{fig:masses}
    Invariant-mass distributions for
    (a) \dmumu, (b) \dee,
    (c) \dmuh1, (d) \deh1,
    (e) \dmuh3, (f) \deh3, (g) \dmue
      candidates with the excluded mass ranges indicated by the gray areas.
    The \Ztautau simulation (red) is normalised to the observed signal.
    The \Z (blue), QCD (brown), and electroweak (magenta) backgrounds are 
    estimated from data. The \ttbar, \VV backgrounds and cross-feed (green) 
    are estimated from simulation (see text) and generally not visible.
  }
\end{minipage}
\end{figure}

After the selections described in the previous Section,
a maximum of one \Ztautau candidate per event is found.
The number of signal candidates is determined from
the number of observed candidates in data
subtracted by the total number of estimated backgrounds.
The results are summarized in \cref{tab:backgrounds}.
The invariant-mass distributions for such
candidates are shown in \cref{fig:masses}, for the seven analysis streams.

A data-driven approach is used to estimate the amount of background from
QCD and \Vj processes.
Same-sign (SS) tau-pair candidates are selected with identical criteria as the signal,
but requiring the tau candidates to have identical electric charge.
From simulation, the SS candidates yield is found to originate mainly from QCD and \Vj processes, 
while the mis-reconstructed \Ztautau process contributes less than 1\%:
$\nss{} = \nss{QCD} + \nss\Vj + \nss\Ztautau$.
The last term originates, for instance, 
from an electron either from a \piz decay or pair-production, 
a single hadron from partially-reconstructed \tauh3, 
3-prong from false combinatorics,
or a muon from a misidentified hadron.
The amount of QCD and \Vj events in the SS dataset is determined by a fit to the
\deltapt distribution, for each analysis stream~\cite{LHCb-PAPER-2012-029}.
In the fit, the QCD distribution templates are taken from an SS QCD-enriched dataset,
obtained by the anti-isolation requirement $\ipthat<0.6$;
the distributions templates for the two \Vj processes (\wjet, \zjet) 
are obtained from simulation and are found to be statistically consistent.
Subsequently, the number of QCD and \Vj background candidates is
computed as $N_\text{QCD}=r_\text{QCD}\cdot\nss{QCD}$, and $N_{\Vj}=r_{\Vj}\cdot\nss{\Vj}$.
The value of $r_{\Vj}$ is obtained from simulation, considering both \W and \Z contributions, 
and varies from $1.05\pm0.08$ for the \ditauee up to $2.37\pm0.30$ for the \ditaumuh1.
The same-sign and opposite-sign QCD-enriched datasets provide the $r_\text{QCD}$ values, 
which are all close to unity, with the exception of $1.30\pm0.05$, obtained for \dmumu.

The \Zll decays ($l=e,\mu$) are a background for all the streams,
except for \ditaumuh3 and \ditaueh3.
The number of \Zll decays contaminating the \dmumu stream is determined
by applying all selection criteria except for the requirement on the dimuon mass: 
this produces a sample with a clear peak at the \Z mass,
as well as an off-shell contribution at lower mass, as shown in \cref{fig:mass_mumu}.
A template distribution obtained from simulation is normalised to the data
in the 80--100\gevcc mass interval.
The fraction of genuine \Ztautau candidates in the normalisation region
is found to be negligible from simulation.
The contribution from \Zmumu decays to the background in the signal region
is inferred from the normalised distribution.
A similar procedure is applied to estimate the \dee background from \Zee decays,
but with the normalisation performed in the \mbox{70--100\gevcc} interval 
to account for the electron momentum resolution
degraded by an incomplete electron bremsstrahlung recovery.
For this process,
1\% of non-\Z background candidates are subtracted from the normalisation region,
as estimated from SS dilepton events.

The process \Zmumu can be observed as a fake \dmuh1 candidate 
when one of the muons is misidentified as a charged hadron.
This background is evaluated by applying the \dmuh1 selection 
but requiring a second identified muon rather than a hadron, 
and scaling by the probability for a muon to be misidentified as a hadron.
The misidentification probability, obtained from simulation and cross-checked
using a tag-and-probe method applied to \Zmumu data 
(requiring an identified muon as a tag, and an oppositely-charged track as a probe),
is of the order of $10^{-3}$ for muons with $\pt<10\gevc$, 
and $10^{-4}$--$10^{-5}$ at larger \pt values.
The uncertainty on the estimation of this background is obtained 
from the lepton misidentification probability uncertainty 
combined with the statistical uncertainty of the dimuon candidates sample. 
A similar procedure allows the estimation of \Zmumu, \Zee backgrounds 
in \dmue, \deh1 streams.

Other background processes are due to diboson decays, \ttbar events,
and \Z decays into \bquark hadrons.
Their contributions are relatively small and obtained from simulation.

Some of the selected tau-pair candidates may not originate from
the stream under study. For instance, a \tauh1 candidate may be
selected from a partially reconstructed \tauh3 candidate.
The fraction of cross-feed candidates is obtained from the \Ztautau simulated sample.
The statistical uncertainty is 1 to 3\%, 
to which a small contribution from the uncertainties 
on the branching fractions of the contaminating streams is added.

\begin{table}
  \newcommand{\zz}{\phantom{000}}
  \newcolumntype{X}{S[table-format=3.1, table-figures-uncertainty=1]}
  \newcolumntype{Y}{S[table-format=4.1, table-figures-uncertainty=1]}
  \centering
  \caption{\label{tab:backgrounds}
    Expected backgrounds yields and total number of candidates observed.
    In the last row the uncertainties are the statistical and systematic contributions combined. 
  }
  \scalebox{0.80}{  
    \begin{tabular}{rXXXXXXY}
      \toprule
        {}            & {\zz\ditaumumu} & {\zz\ditaumuh1} & {\zz\ditaumuh3} & {\zz\ditauee}  & {\zz\ditaueh1} & {\zz\ditaueh3} & {\zz\ditaumue} \\
      \midrule
      \Zll            &  249.7 \pm  8.8 &    1.2 \pm  0.5 &    {\zz ---}    & 420.8 \pm 25.3 &  16.1 \pm  2.2 &   {\zz ---}    &   25.3 \pm  5.4 \\
      QCD             &   50.9 \pm 10.2 &  235.8 \pm 19.3 &   21.2 \pm  5.3 &  42.7 \pm  8.8 & 330.8 \pm 22.8 &  19.4 \pm  5.1 &  160.0 \pm 16.9 \\
      \Vj             &   12.7 \pm  7.4 &  144.2 \pm 43.0 &    5.1 \pm  3.4 &   5.8 \pm  2.7 &  68.3 \pm 19.7 &  10.1 \pm  5.8 &   65.3 \pm 25.7 \\
      \VV             &    0.2 \pm  0.1 &    1.2 \pm  0.2 &    0.2 \pm  0.1 &   0.2 \pm  0.1 &   0.8 \pm  0.1 &   0.2 \pm  0.1 &   10.0 \pm  0.5 \\
      \ttbar          &    1.0 \pm  0.2 &    2.2 \pm  0.2 &    0.6 \pm  0.1 &   0.2 \pm  0.0 &   0.7 \pm  0.1 &   0.1 \pm  0.0 &    5.5 \pm  0.2 \\
      \zbb            &    0.8 \pm  0.4 &    0.3 \pm  0.2 &    0.1 \pm  0.1 &   0.1 \pm  0.1 &   0.3 \pm  0.2 &   0.1 \pm  0.1 &    0.3 \pm  0.2 \\
      Cross-feed      &    4.5 \pm  1.1 &   22.2 \pm  2.5 &   13.9 \pm  2.0 &  13.0 \pm  3.9 &  16.5 \pm  2.4 &   7.3 \pm  1.7 &   52.5 \pm  4.2 \\
      \midrule                                                                                                                                             
      Total bkg.   &  319.9 \pm 12.7 &  407.1 \pm 37.5 &   41.1 \pm  5.3 & 482.7 \pm 24.2 & 433.5 \pm 22.0 &  37.2 \pm  5.8 &  318.9 \pm 23.6 \\
      Observed        &  \num{696}      &  \num{1373}     &  \num{205}      & \num{610}      & \num{861}      &  \num{110}     & \num{1322}      \\
      \Ztautau &  376.1 \pm 29.0 &  965.9 \pm 52.1 &  163.9 \pm 14.2 & 127.3 \pm 32.9 & 427.5 \pm 35.8 &  72.8 \pm 11.1 & 1003.1 \pm 41.8 \\
      \toprule
    \end{tabular}
  }
\end{table}


\section{Cross-section measurement}

The production cross-section of \Z boson to tau-pair
is measured for each analysis stream using
\begin{equation} 
  \sigma_{\ppZtautau} = \frac{
     N_\text{obs} / \erec^\text{obs} 
     - \sum_{\rm k} N_\text{bkg,k} / \erec^\text{bkg,k}
  }{\lum \; \BR \; \eacc \; \esel},
\end{equation} 
where \nobs is the number of observed \Z bosons
and $N_\text{bkg,k}$ is the estimated background from source k.

The total integrated luminosity is denoted by \lum, 
and \BR is the product of the branching fractions 
of the tau lepton pair to decay to the given final state,
with values and uncertainties taken from the world averages~\cite{PDG2016}.
The acceptance factor, \eacc, is needed to normalise the results of each 
analysis stream to the kinematical region $60 < M_{\tau\tau} < 120~\gevcc$, 
$2.0<\eta^\tau<4.5$, and $\pt^\tau > 20~\gevc$,
which allows the comparison with the \Zmumu, \Zee
decay measurements in LHCb~\cite{LHCb-PAPER-2015-049, LHCb-PAPER-2015-003}.
This factor is the fraction of \Ztautau events where the generated \tauon satisfy the chosen
kinematical selections, which also fulfill the fiducial acceptance selection.
The value of \eacc for each stream is obtained from simulation,
using the \texttt{POWHEG-BOX}~\cite{Alioli:2008gx,*Nason:2004rx,*Frixione:2007vw,*Alioli:2010xd} 
at \nlo with PDF \texttt{MSTW08NLO90cl}~\cite{Martin:2009iq},
and \pythia8.175~\cite{Sjostrand:2007gs,*Sjostrand:2006za}.  
The uncertainty on \eacc from the choice of PDF is estimated following the 
procedure explained in Ref.~\cite{Botje:2011sn}.

The event reconstruction and selection efficiencies, \erec and \esel,
as well as their uncertainties,
are estimated from simulation and calibrated using a data-driven method (where applicable)
derived from the method described in
Refs.~\cite{LHCb-PAPER-2012-029,LHCb-PAPER-2015-049, LHCb-PAPER-2015-003}.
The term \erec is the product of the GEC, trigger, tracking and particle identification efficiencies.
The smallest value of \erec is found to be 9\% in the \ditaueh3 stream, 
while the largest value is 65\% for \ditaumumu.
The GEC efficiency is determined from \Zll decays in data 
collected with a relaxed requirement.
The muon and electron trigger efficiencies are evaluated as a function of \Eta and \pt
using a tag-and-probe method applied on \Zll decays.
The tracking efficiency for muons uses a tag-and-probe method from \Zmumu decays in data,
whereas for electrons and charged hadrons simulated samples are used.
The particle identification efficiency is also obtained by a tag-and-probe procedure.
In order to cover the signal \pt spectrum, different data samples are selected:
\Zmumu and $\jpsi\To\mumu$ decays for muons,
\Zee and $\Bp\To\jpsi{(\To\epem)}\Kp$ decays for electrons,
and $\Dstarp\To\Dz{(\To\Km\pip)}\pip$ decays for charged hadrons.

The efficiency of the selection ranges between 20\% for \dee and 50\% for \dmue.
The values are obtained from the simulation.
Corrections at the level of 1\% are inferred by the comparison of the 
selection-variable distributions for \Zmumu decays in data and simulated samples,
which are also added to the systematic uncertainty.

\newcommand{\cscrerrztau}{$\delta\sigma_{\ppZtautau}$\xspace}

\begin{table}
\caption{\label{tab:rerr} 
  Relative uncertainties of the various contributions affecting
  the cross-section measurement, given in percent. 
  The uncertainties are correlated between streams,
  except in rows denoted with $^\dagger$.}
\centering
\begin{tabular}{@{}lrrrrrrr@{}}
\toprule
& \dmumu & \dmuh1 & \dmuh3 & \dee & \deh1 & \deh3 & \dmue\\
\midrule
Tau branching fractions product  & 0.5 & 0.3 & 0.5 &  0.5 & 0.3 &  0.5 & 0.3\\
PDF, acceptance, FSR             & 1.3 & 1.9 & 1.5 &  1.3 & 1.9 &  1.5 & 1.3\\
Reconstruction                   & 2.1 & 3.1 & 5.6 &  4.5 & 5.4 &  7.0 & 2.7\\
Selection                        & 5.0 & 3.5 & 4.7 &  5.7 & 3.5 &  5.1 & 3.9\\
Background estimation$^\dagger$  & 3.4 & 3.9 & 3.2 & 19.0 & 5.2 &  8.0 & 2.4\\
\midrule
Systematic                       & 6.4 & 6.2 & 8.0 & 20.3 & 8.4 & 11.8 & 5.2\\
Statistical$^\dagger$            & 6.9 & 3.8 & 8.1 & 17.6 & 6.6 & 13.1 & 3.4\\
Beam energy                      & 0.2 & 0.2 & 0.2 &  0.2 & 0.2 &  0.2 & 0.2\\
Luminosity                       & 1.2 & 1.2 & 1.2 &  1.2 & 1.2 &  1.2 & 1.2\\
\midrule
Total                            & 9.6 & 7.5 & 11.5 & 27.0 & 10.8 & 17.7 & 6.5\\
\bottomrule
\end{tabular}
\end{table}

A summary of uncertainties is given in \cref{tab:rerr},
with the statistical uncertainty from \nobs obtained assuming Poissonian statistics.
The contribution of the LHC beam energy uncertainty~\cite{PhysRevAccelBeams.20.081003}
is of 0.2\% as studied with the \dynnlo generator~\cite{Catani:2009sm}.
The integrated luminosity is measured using 
van der Meer scans~\cite{vanderMeer:296752}
and beam-gas imaging method~\cite{FerroLuzzi:2005em},
giving a combined uncertainty of 1.2\%~\cite{LHCb-PAPER-2014-047}.

\begin{figure}[!t]
\centering
\includegraphics[width=1.\textwidth]{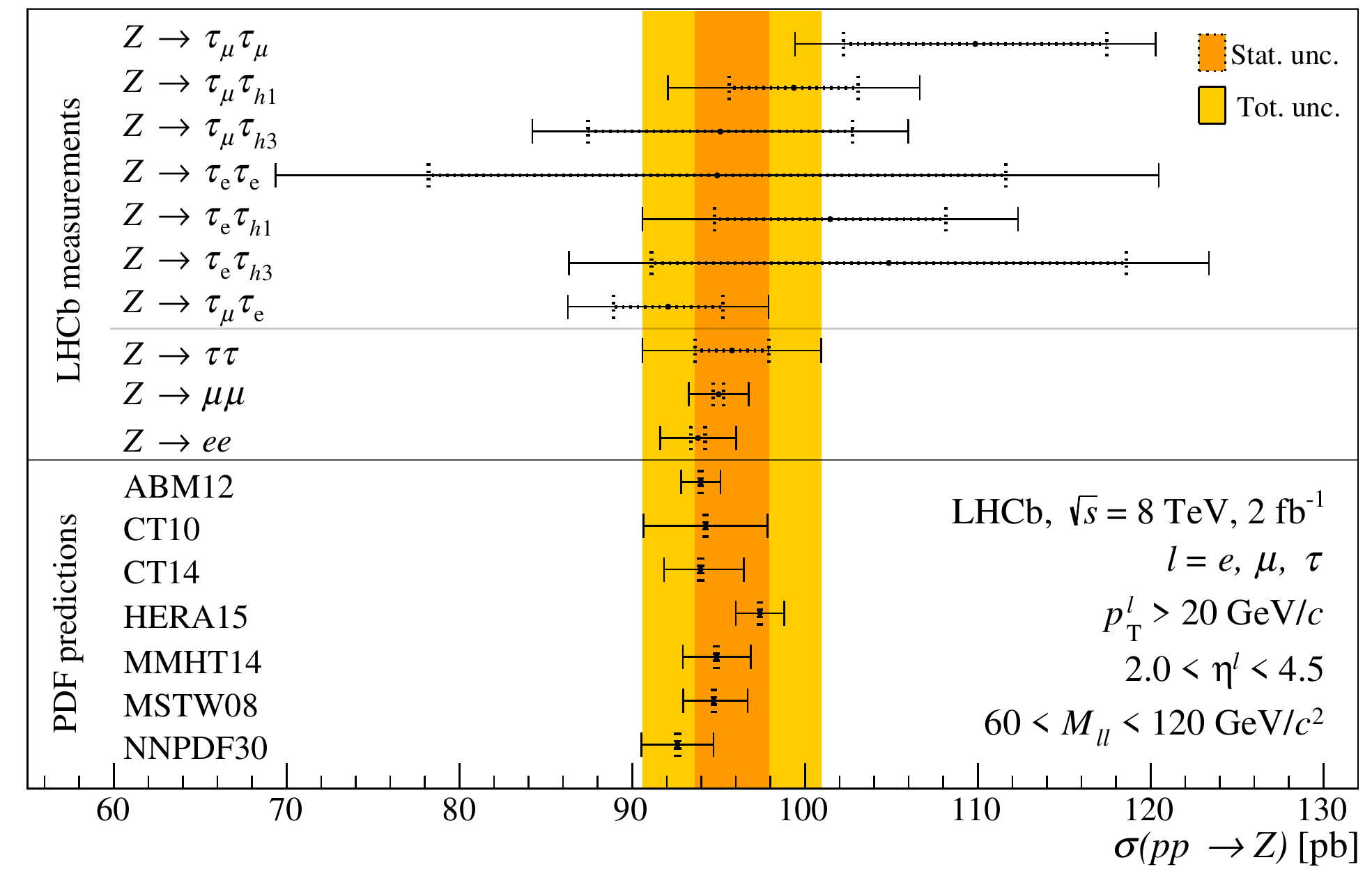}
\vspace*{-20pt}
\caption{\label{fig:measurements}
Summary of the measurements of \Zll production cross-section inside the \lhcb
acceptance region from $pp$ collisions at 8$\mathrm{\,Te\kern -0.1em V}$.
The error bar represents the total uncertainty.
The dotted inner error bar corresponds to the statistical contribution.
The coloured band corresponds to the combined measurement of \Ztautau from this analysis.
The last 7 rows represent the NNLO predictions with different parameterizations
of the PDFs.
}
\end{figure}

The cross-section results compared with the previous \Zmumu and \Zee measurements 
inside the same acceptance region at 8\tev~\cite{LHCb-PAPER-2015-003,LHCb-PAPER-2015-049}, 
are presented in \cref{fig:measurements},
where the region is defined for \Z bosons with an invariant mass between 60 and 120\gevcc
decaying to leptons with $\pt > 20\gevc$ and $2.0<\eta<4.5$.
The predictions from theoretical models are calculated with 
the \fewz~\cite{Li:2012wna,Gavin:2010az} generator at NNLO
for the PDF sets
ABM12~\cite{Alekhin:2013nda},
CT10~\cite{Lai:2010vv},
CT14~\cite{Dulat:2015mca},
HERA15~\cite{Aaron:2009aa},
MSTW08~\cite{Martin:2009iq},
MMHT14~\cite{Harland-Lang:2014zoa}, and
NNPDF30~\cite{Ball:2014uwa}.
A best linear unbiased estimator is used to combine the measurements from all streams
taking into account their correlations,
giving a \chisq per degree of freedom of 0.69 ($p$-value of 0.658).
The combined cross-section is
\begin{equation*}
\sigma_{\ppZtautau} = 95.8 \pm 2.1 \pm 4.6 \pm 0.2 \pm 1.1 \pb \,,
\end{equation*} 
where the uncertainties are statistical, systematic, 
due to the LHC beam energy uncertainty, 
and to the integrated luminosity uncertainty, respectively.

Lepton universality is tested from the cross-section ratios
\cite{LHCb-PAPER-2015-049, LHCb-PAPER-2015-003}
\begin{equation*} 
\frac{\sigma^{8\tev}_{\ppZtautau} }{\sigma^{8\tev}_{\ppZmumu} } = 1.01 \pm 0.05
\quad , \quad
\frac{\sigma^{8\tev}_{\ppZtautau} }{\sigma^{8\tev}_{\ppZee} } = 1.02 \pm 0.06
\,,
\end{equation*} 
where the uncertainties due to the LHC beam energy and to the
integrated luminosity are assumed to
be fully correlated as the analyses share the same dataset,
whilst the statistical and systematic uncertainties are
assumed to be uncorrelated.

\section{Conclusion}

A measurement of \Ztautau production cross-section in $pp$ collisions at \sqs = 8\tev
inside \lhcb fiducial acceptance region is reported, where the region
is defined as a tau-pair of invariant mass between 60 and 120\gevcc, 
with the tau leptons having a transverse momentum greater than 20\gevc, 
and pseudorapidity between 2.0 and 4.5.

The reconstruction of tau-pair candidates is performed in both leptonic and
hadronic decay modes of the tau lepton, requiring at least one leptonic
mode for the tau-pair combination.
The backgrounds to \Ztautau are mainly from QCD and \W/\Z+jets and
are estimated with a data-driven method.

The production cross-section with all uncertainties summed in quadrature 
yields \num{95.8 +- 5.2}\pb, in agreement with the SM prediction. 
The results are consistent with the \Zee and \Zmumu cross-sections measured at LHCb.
They are compatible with LU at the level of 6\%.

\section*{Acknowledgements}
%
%
\noindent We express our gratitude to our colleagues in the CERN
accelerator departments for the excellent performance of the LHC. We
thank the technical and administrative staff at the LHCb
institutes. We acknowledge support from CERN and from the national
agencies: CAPES, CNPq, FAPERJ and FINEP (Brazil); MOST and NSFC
(China); CNRS/IN2P3 (France); BMBF, DFG and MPG (Germany); INFN
(Italy); NWO (Netherlands); MNiSW and NCN (Poland); MEN/IFA
(Romania); MinES and FASO (Russia); MinECo (Spain); SNSF and SER
(Switzerland); NASU (Ukraine); STFC (United Kingdom); NSF (USA).  We
acknowledge the computing resources that are provided by CERN, IN2P3
(France), KIT and DESY (Germany), INFN (Italy), SURF (Netherlands),
PIC (Spain), GridPP (United Kingdom), RRCKI and Yandex
LLC (Russia), CSCS (Switzerland), IFIN-HH (Romania), CBPF (Brazil),
PL-GRID (Poland) and OSC (USA). We are indebted to the communities
behind the multiple open-source software packages on which we depend.
Individual groups or members have received support from AvH Foundation
(Germany), EPLANET, Marie Sk\l{}odowska-Curie Actions and ERC
(European Union), ANR, Labex P2IO and OCEVU, and R\'{e}gion
Auvergne-Rh\^{o}ne-Alpes (France), Key Research Program of Frontier
Sciences of CAS, CAS PIFI, and the Thousand Talents Program (China),
RFBR, RSF and Yandex LLC (Russia), GVA, XuntaGal and GENCAT (Spain),
Herchel Smith Fund, the Royal Society, the English-Speaking Union and
the Leverhulme Trust (United Kingdom).

\addcontentsline{toc}{section}{References}
\setboolean{inbibliography}{true}
\bibliographystyle{LHCb}
\bibliography{main,LHCb-PAPER,LHCb-CONF,LHCb-DP,LHCb-TDR,manual}

\newpage


 
\newpage
\centerline{\large\bf LHCb collaboration}
\begin{flushleft}
\small
R.~Aaij$^{27}$,
B.~Adeva$^{41}$,
M.~Adinolfi$^{48}$,
C.A.~Aidala$^{73}$,
Z.~Ajaltouni$^{5}$,
S.~Akar$^{59}$,
P.~Albicocco$^{18}$,
J.~Albrecht$^{10}$,
F.~Alessio$^{42}$,
M.~Alexander$^{53}$
% ...

{\footnotesize \it
% ...
$ ^{5}$Clermont Universit{\'e}, Universit{\'e} Blaise Pascal, CNRS/IN2P3, LPC, Clermont-Ferrand, France\\
% ...
$ ^{10}$Fakult{\"a}t Physik, Technische Universit{\"a}t Dortmund, Dortmund, Germany\\
% ...
$ ^{18}$INFN Laboratori Nazionali di Frascati, Frascati, Italy\\
% ...
$ ^{27}$Nikhef National Institute for Subatomic Physics, Amsterdam, Netherlands\\
% ...
$ ^{41}$Instituto Galego de F{\'\i}sica de Altas Enerx{\'\i}as (IGFAE), Universidade de Santiago de Compostela, Santiago de Compostela, Spain\\
$ ^{42}$European Organization for Nuclear Research (CERN), Geneva, Switzerland\\
% ...
$ ^{48}$H.H. Wills Physics Laboratory, University of Bristol, Bristol, United Kingdom\\
% ...
$ ^{53}$School of Physics and Astronomy, University of Glasgow, Glasgow, United Kingdom\\
% ...
$ ^{59}$University of Cincinnati, Cincinnati, OH, United States\\
% ...
$ ^{61}$Syracuse University, Syracuse, NY, United States\\
% ...
$ ^{73}$University of Michigan, Ann Arbor, United States, associated to $^{61}$\\
% ...
}
\end{flushleft}

\end{document}